\documentclass[12pt]{smdqart}
\usepackage{amsfonts}
\usepackage{amssymb}
\usepackage{graphicx}
\usepackage{epsfig}
\usepackage{color}

\begin{document}

\title{Minding impacting events in a model of stochastic variance}
\author{S M Duarte Queir\'{o}s $^{1, \ast}$, E M F Curado $^2$ and F D Nobre $^2$}

\address{$^1$ Centro de F\'{\i}sica do Porto, Rua do Campo Alegre 687, 4169-007, Porto, Portugal}

\address{$^2$ Centro Brasileiro de Pesquisas F\'{\i}sicas and National Institute of Science and Technology for Complex Systems,
Rua Dr Xavier Sigaud 150, 22290-180, Rio de Janeiro - RJ, Brazil \\}

\address{$^{\ast }$ Present address: Istituto dei Sistemi Complessi - CNR, Via dei Taurini, 19, 00185 Roma, Italy}

\ead{\mailto{sdqueiro@gmail.com}}

\begin{abstract}
We introduce a generalisation of the well-known ARCH process, widely used for generating uncorrelated stochastic time series with long-term non-Gaussian distributions and long-lasting correlations in the (instantaneous) standard deviation exhibiting a clustering profile.
Specifically, inspired by the fact that in a variety of systems impacting events are hardly forgot, we split the process into two different regimes:
a first one for regular periods where the average volatility of the fluctuations within a certain period of time $W$ is below a certain
threshold, $\phi $, and another one when the local standard deviation outnumbers $\phi $.
In the former situation we use standard rules for heteroscedastic processes whereas in the latter case the system starts
recalling past values that surpassed the threshold. Our results show that for appropriate
parameter values the model is able to provide fat tailed probability density functions and strong persistence of the instantaneous variance
characterised by large values of the Hurst exponent ($H > 0.8$), which are ubiquitous features in complex systems.
\end{abstract}


\keyw{Heteroscedastic processes, Fat-tail distributions, Perpetual memory}



\maketitle

\section{Introduction}
For the last years the physical community has broaden its subject goals to
matters that some decades ago were too distant from the classical topics of
Physics. Despite being apparently at odds with the standard motivations of Physics, this new trend has given an invaluable contribution
toward a more connected way of making Science, thus leading to a better
understanding of the world surrounding us~\cite{gell-mann}. Within this context, the major contribution of
physicists is perhaps the quantitative procedure, reminiscent of experimental physics, in which a model is
proposed after a series of studies that pave the way to a reliable theory. This path has resulted in a series of findings
which have helped such diverse fields as physiology, sociology and
economics, among many others~\cite{santafe1,santafe2,complex}. Along these findings, one can mention
the determination of non-Gaussian distributions and long-lasting (power-law like)
correlations~\cite{bouchaudpotters,mantegnastanley,voit}. Actually, by
changing the observable, the conjunction of the two previous empirical
verifications is quite omnipresent. For this reason and regardless the realm of the problem very similar models
have been applied with particular notoriety to discrete stochastic processes of time-dependent variance based on
autoregressive conditional heteroscedastic models~\cite{andersen}. That is to say, most of these
models are devised taking basically into account the general features one
aims at reproducing, rather than putting in elements that represent the
idiosyncracies of the system one is surveying. For instance, many of the proposals
cast aside the cognitive essence prevailing on many of these systems, when it is well known that in real situations this represents a
key element of the process~\cite{cognitive}. On the other hand, intending to describe
long-lasting correlations, long-lasting memories are usually introduced thus neglecting
the fact that we do not traditionally keep in mind every happening. As a simple example, we are skilled at
remembering quotidian events for some period. However, we will discard that information
as time goes by, unless the specific deed either created an impact on us or has to do with something that has really
touched us somehow. In this case, it is likely that the fact will be remembered forever and called back in
similar or related conditions, which many times lead to a collective memory effect~\cite{collective}.

In this work, we make use of the celebrated heteroscedastic model, the $ARCH
$ process~\cite{engle} and modify it by pitching at accommodating cognitive traits that lead to different behavior for periods of high
agitation or impact. Particularly, we want to stress on the fact that people
tend to recall important periods, no matter when they took place.
To that end, we introduce a measure of the local volatility, as well
as a volatility threshold, so that the system changes from a normal dynamics,
in which it uses the previous values of the variable to determine its next value, to a situation in which it recalls the past
and compares the current state with previous states of high volatility, even if this past is far.

\subsection{Standard models of heteroscedasticity}
The Engle's formulation of an autoregressive conditional heteroscedastic ($%
ARCH$) time series~\cite{engle} represents one of the simplest and effectual models
in Economics and Finance, for which he was laureated the
Nobel Memorial Prize in Economical Sciences in 2003~\cite{nobel}.
Explicitly, the $ARCH$ corresponds to a discrete time, $t$, process associated with a
variable, $z_{t}$,
\begin{equation}
z_{t}=\sigma _{t}\ \omega _{t},  \label{arch-def}
\end{equation}%
with $\omega _{t}$ being an independent and identically distributed random
variable with zero mean and standard deviation equal to one. The quantity $\sigma _{t}$ represents the
time-dependent standard deviation, which we will henceforth name \textit{\
instantaneous volatility} for mere historical reasons. Traditionally, a Gaussian
is assigned to the random variable $\omega _t$, but other
distributions, namely the truncated $\alpha $-stable L\'{e}vy distribution
and the $q$-Gaussian (Student-$t$) have been successfully introduced as
well~\cite{noise-gen,smdqct}. In his seminal paper, Engle suggested that the values of~$\sigma
_{t}^{2}$ could be obtained from a linear function of past squared values of
$z_{t}$,
\begin{equation}
\sigma _{t}^{2}=a+\sum\limits_{i=1}^{s}b_{i}\ z_{t-i}^{2},\qquad \left(
a,b_{i}\geq 0\right) .  \label{arch-vol}
\end{equation}%
In financial practice, \textit{viz.,} price fluctuations modelling, the case $%
s=1$ ($b_{1}\equiv b$) represents the very most studied and applied of all
the $ARCH\left( s\right) $-like processes. The model has been often applied in cases where it is assumed that
the variance of the observable (or its fluctuation) is a function of the magnitudes of the previous occurrences.
In a financial perspective, Engle's proposal has been associated with the relation between the market activity and the deviations
from the normal level of volatility $a$, and the previous price fluctuations making use of the impact function~\cite{andersen}.
Alternatively, recent studies convey the thesis that leverage can be responsible for the volatility clustering and fat
tails in finance~\cite{thurnerfarmer}. Nonetheless, the heteroscedastic $ARCH$-like processes has been repeatedly used as a forecasting method. In other words, one makes use
of the magnitude of previous events in order to indicate (or at least to bound) the upcoming event (see e.g.~\cite{forecast1,forecast2}).
In respect of its statistical features, although the time series is completely uncorrelated, $%
\left\langle z_{t}\ z_{t^{\prime }}\right\rangle \sim \delta _{t\,t^{\prime
}}$, it can be easily verified that the covariance $\left\langle \left\vert
z_{t}\right\vert \ \left\vert z_{t^{\prime }}\right\vert \right\rangle $ is
not proportional to $\delta _{t\,t^{\prime }}$. As a matter of fact, for $%
s=1 $, it is provable that $\left\langle z_{t}^{2}\ z_{t^{\prime
}}^{2}\right\rangle $ decays according to an exponential law with a
characteristic time $\tau \equiv \left\vert \ln b\right\vert ^{-1}$. This dependence
does not reproduce most of the empirical evidences, particularly those
bearing on price fluctuations studies. In addition, the introduction of a
large value of $s$ used to give rise to implementation problems~\cite{boller}.
Expressly, large values of $s$ augment the difficulty of finding the
appropriate set of parameters $\left\{ b_{i}\right\} $ for the problem under
study as it corresponds to the evaluation of a large number of fitting
parameters. Aiming to solve this short-coming of the original $ARCH\left(
1\right) $ process, the $GARCH\left( s,r\right) $ process was introduced~%
\cite{granger} (where $G$ stands for generalised), with Eq. (\ref{arch-vol})
being replaced by,
\begin{equation}
\sigma _{t}^{2}=a+\sum_{i=1}^{s}b_{i}\
z_{t-i}^{2}+\sum\limits_{i=1}^{r}c_{i}\ \sigma _{t-i}^{2}\qquad \left(
a,b_{i},c_{i}\geq 0\right) .  \label{garch}
\end{equation}%
In spite of the fact that the condition, $b+c<1$, guarantees that the $%
GARCH\left( 1,1\right) $ process exactly corresponds to an infinite-order $%
ARCH$ process, an exponential decay for $\left\langle z_{t}^{2}\
z_{t^{\prime }}^{2}\right\rangle $, with $\tau \equiv \left\vert \ln \left(
b+c\right) \right\vert ^{-1}$ is found.

Although the instantaneous volatility is time dependent, the $ARCH(1) $ process
is actually stationary with the \textit{stationary variance} given by,
\begin{equation}
\left\langle \sigma ^{2}\right\rangle =\widehat{\sigma ^{2}}=\frac{a}{1-b}%
,\qquad (b<1),  \label{statvar}
\end{equation}
(herein $\left\langle \ldots \right\rangle $ represents averages over
samples at a specified time and $\widehat{\ldots }$ denotes averages over time in a
single sample). Moreover, it presents a stationary probability density
function (PDF), $P\left( z\right) $, with a kurtosis larger than the
kurtosis of distribution $P(\omega )$. Namely, the fourth-order moment is,
\[
\left\langle z^{4}\right\rangle =a^{2}\left\langle \omega ^{4}\right\rangle
\frac{1+b}{\left( 1-b\right) \left( 1-b^{2}\left\langle \sigma
^{4}\right\rangle \right) }.
\]%
This kurtosis excess is precisely the outcome of the dependence of $\sigma _{t}$ on the time (through $z$).
Correspondingly, when $b=0$, the process is reduced
to generating a signal with the same PDF of $\omega $, but with a standard
variation equal to $\sqrt{a}$. At this point, it is convenient to say that, for the time being and despite several efforts, there are only analytical
expressions describing the tail behaviour of $P(z)$ or the continuous-time
approximation of the $ARCH$(1) process with the full analytical formula still
unknown~\cite{smdqct,embrechts}.

In order to cope with the long-lasting correlations and other features such
as the asymmetry of the distribution and the leverage effect, different versions
of the $ARCH$ process have been proposed~\cite{andersen,boller}.
To the best of our knowledge, every of them solve the issue
of the long-lasting correlations of the volatility by way of introducing an
eternal dependence on $z_{i}^{2}$ in Eq. (\ref{arch-vol}), $b_{i}\equiv b\,%
\mathcal{K}\left( i\right) $, with $\mathcal{K}\left( .\right) $
representing a slowly decaying function~\cite{goug,pier}. Most of these
generalisations can be encompassed within the fractionally integrated class
of $ARCH$ processes, the $FIARCH$ \cite{fiarch,porto,q-archepl}.
The idea supporting the introduction of a power-law for the functional form
of $\mathcal{K}\left( .\right) $ is generally based on the assumption that the
agents in the market make use of exponential functions $\mathcal{K}\left(
.\right) $ with a broad distribution of relaxation times related to
different investment horizons \cite{dacorogna,bouchaudintro}. This type of model has achieved a huge
popularity in the replication of non-Gaussian time series in several areas,
such as biomedicine, climate, engineering, and physics (a few examples can be
found in~\cite{climate,campbell,martin-guerrero,gronke,reynolds,beckprl}).

As described above, the statistical features of the macroscopic
observables are the result of the nature of the interactions between the
microscopic elements of the system and the relation between microscopic as well as the macroscopic observables. In the case of the ``financial''
$ARCH$ process, it was held that $z_{i}^{2}$ bears upon the impact of the
price fluctuations on the trading activity.
On the one hand, it is understood that the impact of the price fluctuations
(or trading activity) on the volatility does not merely come from recent
price fluctuations and it does actually involve past price fluctuations.
In finance, upgraded versions of heteroscedasticity models use multi-scaling, \textit{i.e.}, it is assumed that the
price will evolve by modulating the volatility
according to the volatility over different scales (days, weeks, months,
years, etc.)~\cite{zumbach} in order to smooth their possible misjudgement
about the volatility. However, in practice, these models do not differ
much from $FIARCH$-like proposals at the level of the results we are
pointing at. Alternately, it is worthwhile to look upon the $ARCH$
proposal as a mechanism of forecast~\cite{forecast1,forecast2}. In this way, the simplest approach, the
$ARCH(1)$, represents an attempt to foresee future values just taking into account recent observations,
whereas models like the $FIARCH$ bear in mind all the history weighting each
past-value according to some kernel functional.

\subsection{Minding impacting events}

In our case, we want to emphasise the fact that people tend to recall
periods of high volatility (\textit{i.e.}, impact) in the system, no matter
when they took place, by changing the surrounding conditions as agent-based models suggested~\cite{lux,giardina}.
Hence, we introduce a measure of the local volatility,%
\begin{equation}
v_{t}=\frac{1}{W}\sum\limits_{i=0}^{W-1}z_{t-i}^{2},
\end{equation}%
and a threshold, $\phi $, so that instead of Eq.~(\ref{arch-vol}), the
updating of $\sigma _{t}^{2}$ goes as follows:%
\begin{equation}
\sigma _{t}^{2}=\left\{
\begin{array}{ccc}
a+\sum\limits_{i=1}^{t}b_{i}\,\ z_{t-i}^{2} & \mathrm{if} & v_{t-1}<\phi ,
\\
&  &  \\
a+\sum\limits_{i=1}^{t}b_{i}^{\prime }\ z_{t-i}^{2} & \mathrm{if} &
v_{t-1}\geq \phi ,%
\end{array}%
\right.  \label{itarchvol}
\end{equation}%
where $b_{i}=b\,\mathcal{K}\left( i\right) =b\,\exp \left[ -\,i\,/\,\tau %
\right] $~\cite{exponential}. Therefore, if we assume the financial market perspective, we
are implicitly presuming that the characteristic time, $\tau $, is Dirac
delta or at least narrow distributed, so that the exponential functional
is a valid approximation. This approach is confirmed by recent heuristic studies in
which it has been verified that the largest stake of the market
capitalisation is managed by a small number of companies that apply very
similar strategies~\cite{borlandsp}. With the second branch equation we intend
to highlight the difference in behaviour of the \textquotedblleft
normal\textquotedblright\ periods of trading and the periods of significant
volatility, in which the future depends on the spells of
significant volatility in the past as well. The values $b_{i}^{\prime }$ are defined
as,%
\begin{equation}
b_{i}^{\prime }=b\,p_{i}\,\Theta \left[ v_{t-i}-\phi \right] ,
\end{equation}%
with $\Theta \left[ \ldots \right] $ being the Heaviside function and $p_{i}$ is a factor that represents a measure of the
similarity (in the volatility space) between the windows of size $W$ with
upper limits at $z_{t}$ and $z_{t-W+1}$, respectively. Analytically, this
is equivalent to mapping segments in the form $\left\{ z_{i}^{2},\ldots
,z_{i-W+1}^{2}\right\} $ into vectors in $\mathbb{R}_{0}^{+\,W}$ and
afterward computing a normalised internal product-like weight,%
\begin{equation}
p_{i}=\frac{1}{\mathcal{N}}\sum\limits_{j=1}^{W}z_{t-j}^{2}\,z_{i-j}^{2},
\label{dotproduct}
\end{equation}%
where, for the sake of simplicity, we set aside the time dependence of $p _i$ and $b_{i}^{\prime }$
in the equations, while $\mathcal{N}$ represents the normalisation factor such that $\sum_{i}p_{i}=1$ for all $i$ (with fixed $t$).

We are therefore dealing with a model characterised by 5 parameters, namely: $a$ (the normal level of volatility) and
$b$ (the impact of the observable in the volatility), which were both first introduced by Engle in~\cite{engle};
$\tau $, put forward in exponential models~\cite{exponential}; and two new parameters $W$ (representing the volatility spell) and $\phi $ that we will
reduce to a single extra parameter.
If we think of trading activities, our proposal introduces a key parameter, the volatility threshold, $\phi $,
which signals a change in behaviour of the agents in the market. At
present, significant stake of the trading in financial markets is dominated by
short-term positions and thus a good part of the dynamics of price
fluctuations can be described by Eq.~(\ref{arch-vol}), or by functions with
an exponential kernel. As soon as the market fluctuates excessively, \textit{i.e.}, the
volatility soars beyond the threshold, the market changes its trading
dynamics. The main forecast references are obviously the periods where
the volatility has reached high levels and afterward, the periods of those which
are most similar; this is the rationale described by our Eq.~(\ref{dotproduct}).
Thence, our proposal is nothing but the use of simple mechanisms that
in a coarse-grained way master a good part of our decisions.

\section{Results}

\subsection{General results}

In this section we present the results obtained by the numerical
implementation of the model. For comparison, we will use the results of a prior model that can be
enclosed in the class of $FIARCH$ processes~\cite{q-archepl}. There, the adjustment of
the parameters comes from the delicate balance between the parameter $b$,
which is responsible for introducing deviations of the volatility from its normal level $a$,
and the parameter controlling the memory. On the one
hand, large memory has the inconvenient effect of turning constant the
instantaneous volatility, so that after a seemly number of time steps the value of $\sigma $ becomes constant,
hence leading to a Gaussian (or close to it) distribution of the variable $z$, independently of how large $b$ is. On the other
hand, short memory is unable to introduce long-range correlations in the
volatility, although it enhances larger values of kurtosis excess. The model we
introduce herein is rather more complex. In order to deal with the change of
regime, we define a parameter establishing this alteration and we need
to specify $W$ and $\tau $. Henceforth, we have assumed $W=\tau $, which is very reasonable as it imposes that the volatility and the
time scale that the agents in the market use to assess the evolution of the
observable are the same. In order to speed up our numerical
implementation, we have imposed a cut-off of $10W$ in the computation of the
first line in Eq. (\ref{itarchvol}). This approximation turns the numerical
procedure much lighter with a negligible effect because the influence of the
discarded past is not much relevant in numerical terms (within standard numerical implementation error). In all of our realisations, we have
used a normalised level of expected volatility, $a=1$, and we have defined
the volatility threshold in units of $a/\left( 1-b\right) $, following a
stationary approach, as well.

We have adjusted the probability distributions of $z$ by means of the distribution,%
\begin{equation}
P\left( z\right) =\mathcal{Z}^{-1}\left( 1+B\,z^{2\,\nu }\right) ^{\frac{1%
}{1-q^{\prime }}},  \label{pdfz}
\end{equation}%
the behaviour of which follows a power-law distribution for large $|z|$ with
an exponent equal to $\frac{2\,\nu }{q^{\prime }-1}$ and where (using Ref.~\cite{fouriertransforms},~sec.~ 3.194),
\begin{equation}
\int z^{n}(1+B\,z^{2\,\nu })^{\frac{1}{1-q^{\prime }}}\,dz=\frac{1+\left( -1\right)
^{n}}{1+n}\frac{\Gamma \left[ \frac{2\,\nu +n+1}{2\,\nu }\right]
\Gamma \left[ \frac{2\,\nu +\left( n+1\right) \left( 1-q^{\prime }\right) }{%
2\,\nu \left( q^{\prime }-1\right) }\right] }{B^{\frac{1+n}{2\,\nu }}\,\Gamma %
\left[ \frac{1}{q^{\prime }-1}\right] },
\end{equation}
$\left( \frac{2\,\nu }{q^{\prime }-1}>1+n\right) ,$ and $\mathcal{Z}$ represents the
previous integral with $n=0$. The fittings for the probability density
distribution~(\ref{pdfz}) were obtained using non-linear and maximum
log-likelihood numerical procedures and the tail exponents double-checked with the
value given by the Hill estimator~\cite{hill}. As a matter of fact, values of $%
\nu $ different from $1$ have only been perceived for large
values of $b$ and small values of $\phi $ (slightly larger) or large values of $\phi $ (slightly smaller).
For $\nu =1$ and $q^{\prime } \neq 1$, the PDF corresponds to a $q^{\prime }$-Gaussian distribution (or Student-$t$ distribution)~\cite{ct-bjp} and when $q^{\prime } = 1$ we have either the Gaussian ($\nu =1$) or the stretched distribution ($\nu \neq 1$). Since that in the majority of
the applications one is interested in the tail behaviour, we have opted for
following the same approach by defining the tail index as,
\begin{equation}
\frac{2}{q-1}=\frac{2 \nu}{q^{\prime } -1 } \, \Leftrightarrow \, q=\nu ^{-1}\left( q^{\prime }+\nu -1\right) .
\end{equation}
In spite of the fact that other functional forms could have been used, we have decided on Eq.~(\ref{pdfz}) because of its statistical relevance and simplicity (in comparison with other candidates involving special functions, namely the hypergeometric).
Moreover, the $q$-Gaussian ($t$-Student) is intimately associated with the long-term
distribution of heteroscedastic variables since it results in the exact distribution
when the volatility follows an inverse-Gamma distribution~\cite{beckprl,beckcohen,smdqEPJB}.

Concerning the persistence of the volatility, we have settled on the
Detrended Fluctuation Analysis (DFA)~\cite{dfa}, which
describes the scaling of a
fluctuation function related to the average aggregated variance over
segments of a time series of size $\ell $,%
\begin{equation}
F\left( \ell \right) \sim \ell ^{H},
\end{equation}%
where $H$ is the Hurst exponent. Although it has been shown that
Fluctuation Analysis methods can introduce meaningful errors in the L\'{e}vy
regime~\cite{dfacheck}, we have verified that for our case,
which stands
within the finite second-order moment domain, the results of DFA are so reliable as other scaling methods.

\begin{table}
\begin{center}
\begin{tabular}{ccccccc}
\cline{1-3}\cline{5-7}
$W$ & $\phi $ & $\textrm{P}_{KS}^{*}$ &  & $W$ & $\phi $ & $\textrm{P}_{KS}^{*}$ \\
\cline{1-3}\cline{5-7}
\multicolumn{1}{l}{$10$} & \multicolumn{1}{l}{$0.25$} & \multicolumn{1}{l}{$%
0.9997$} & \multicolumn{1}{l}{} & \multicolumn{1}{l}{$75$} &
\multicolumn{1}{l}{$0.25$} & \multicolumn{1}{l}{$0.9865$} \\
\multicolumn{1}{l}{} & \multicolumn{1}{l}{$0.5$} & \multicolumn{1}{l}{$0.9998
$} & \multicolumn{1}{l}{} & \multicolumn{1}{l}{} & \multicolumn{1}{l}{$0.5$}
& \multicolumn{1}{l}{$0.9898$} \\
\multicolumn{1}{l}{} & \multicolumn{1}{l}{$0.6$} & \multicolumn{1}{l}{$0.9998
$} & \multicolumn{1}{l}{} & \multicolumn{1}{l}{} & \multicolumn{1}{l}{$0.6$}
& \multicolumn{1}{l}{$0.9902$} \\
\multicolumn{1}{l}{} & \multicolumn{1}{l}{$0.75$} & \multicolumn{1}{l}{$%
0.9998$} & \multicolumn{1}{l}{} & \multicolumn{1}{l}{} & \multicolumn{1}{l}{$%
0.75$} & \multicolumn{1}{l}{$0.9908$} \\
\multicolumn{1}{l}{} & \multicolumn{1}{l}{$1.25$} & \multicolumn{1}{l}{$%
0.9999$} & \multicolumn{1}{l}{} & \multicolumn{1}{l}{} & \multicolumn{1}{l}{$%
1.25$} & \multicolumn{1}{l}{$0.9918$} \\
\multicolumn{1}{l}{} & \multicolumn{1}{l}{$2.5$} & \multicolumn{1}{l}{$0.9999
$} & \multicolumn{1}{l}{} & \multicolumn{1}{l}{} & \multicolumn{1}{l}{$2.5$}
& \multicolumn{1}{l}{$0.9925$} \\
\multicolumn{1}{l}{} & \multicolumn{1}{l}{$5$} & \multicolumn{1}{l}{$1$} &
\multicolumn{1}{l}{} & \multicolumn{1}{l}{} & \multicolumn{1}{l}{$5$} &
\multicolumn{1}{l}{$0.9943$} \\
\multicolumn{1}{l}{$25$} & \multicolumn{1}{l}{$0.25$} & \multicolumn{1}{l}{$%
0.9985$} & \multicolumn{1}{l}{} & \multicolumn{1}{l}{$125$} &
\multicolumn{1}{l}{$0.25$} & \multicolumn{1}{l}{$0.9749$} \\
\multicolumn{1}{l}{} & \multicolumn{1}{l}{$0.5$} & \multicolumn{1}{l}{$0.9989
$} & \multicolumn{1}{l}{} & \multicolumn{1}{l}{} & \multicolumn{1}{l}{$0.5$}
& \multicolumn{1}{l}{$0.9761$} \\
\multicolumn{1}{l}{} & \multicolumn{1}{l}{$0.6$} & \multicolumn{1}{l}{$0.999$%
} & \multicolumn{1}{l}{} & \multicolumn{1}{l}{} & \multicolumn{1}{l}{$0.6$}
& \multicolumn{1}{l}{$0.976$} \\
\multicolumn{1}{l}{} & \multicolumn{1}{l}{$0.75$} & \multicolumn{1}{l}{$%
0.9991$} & \multicolumn{1}{l}{} & \multicolumn{1}{l}{} & \multicolumn{1}{l}{$%
0.75$} & \multicolumn{1}{l}{$0.9767$} \\
\multicolumn{1}{l}{} & \multicolumn{1}{l}{$1.25$} & \multicolumn{1}{l}{$%
0.9992$} & \multicolumn{1}{l}{} & \multicolumn{1}{l}{} & \multicolumn{1}{l}{$%
1.25$} & \multicolumn{1}{l}{$0.9780$} \\
\multicolumn{1}{l}{} & \multicolumn{1}{l}{$2.5$} & \multicolumn{1}{l}{$0.9994
$} & \multicolumn{1}{l}{} & \multicolumn{1}{l}{} & \multicolumn{1}{l}{$2.5$}
& \multicolumn{1}{l}{$0.9817$} \\
\multicolumn{1}{l}{} & \multicolumn{1}{l}{$5$} & \multicolumn{1}{l}{$0.9996$}
& \multicolumn{1}{l}{} & \multicolumn{1}{l}{} & \multicolumn{1}{l}{$5$} &
\multicolumn{1}{l}{$0.9870$} \\ \cline{1-3}\cline{5-7}
\end{tabular}
\end{center}
\caption{Critical values $\textrm{P}_{KS}^{*}= 1- \alpha _{crit}$ from the Kolmogorov-Smirnov test
 for typical pairs $(W,\phi)$ used for adjustments.}
\label{tabks}
\end{table}

\begin{figure}[tbh]
\begin{center}
\includegraphics[width=0.45\columnwidth,angle=0]{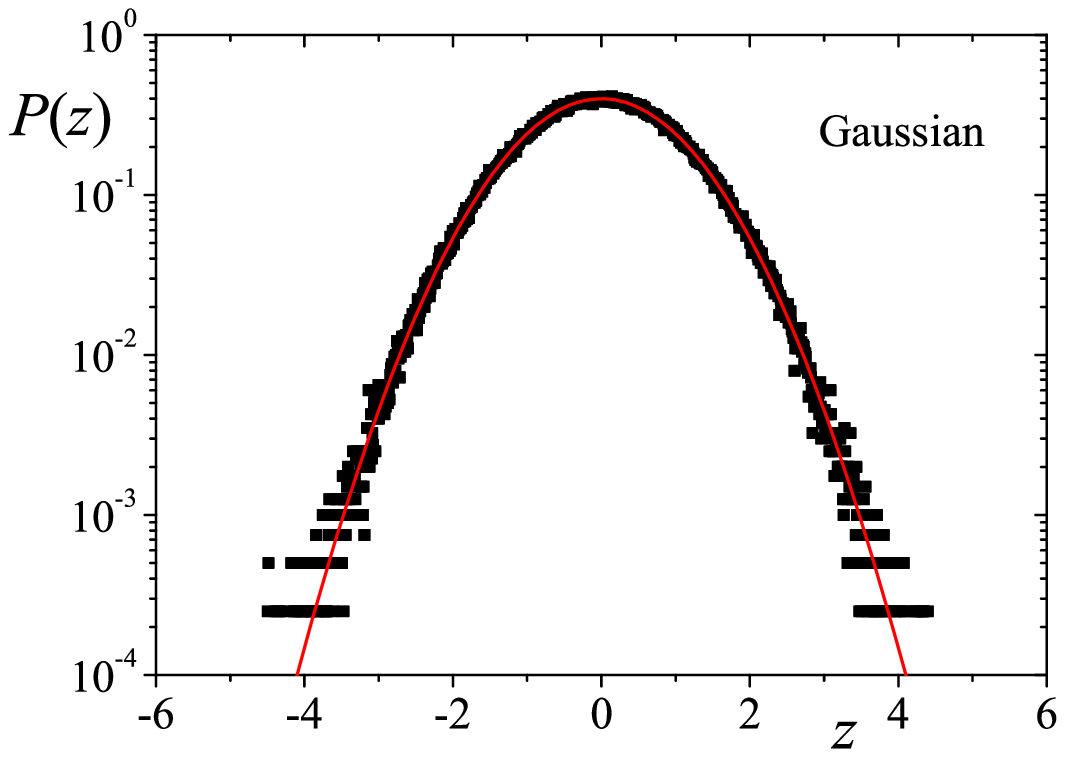} %
\includegraphics[width=0.45\columnwidth,angle=0]{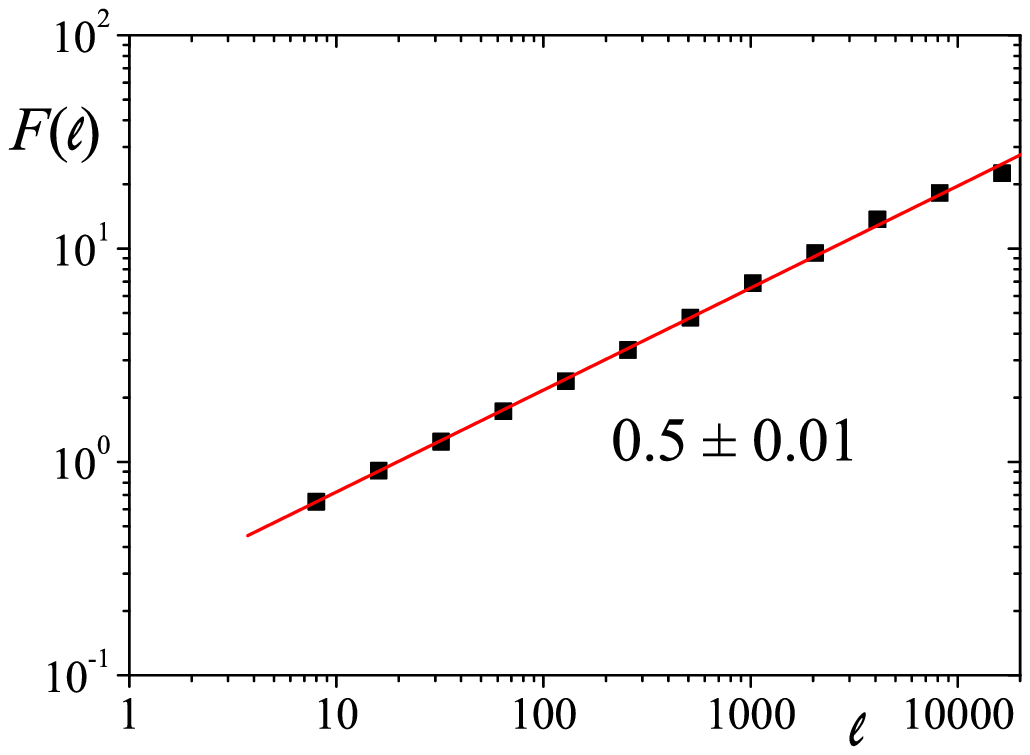} %
\includegraphics[width=0.45\columnwidth,angle=0]{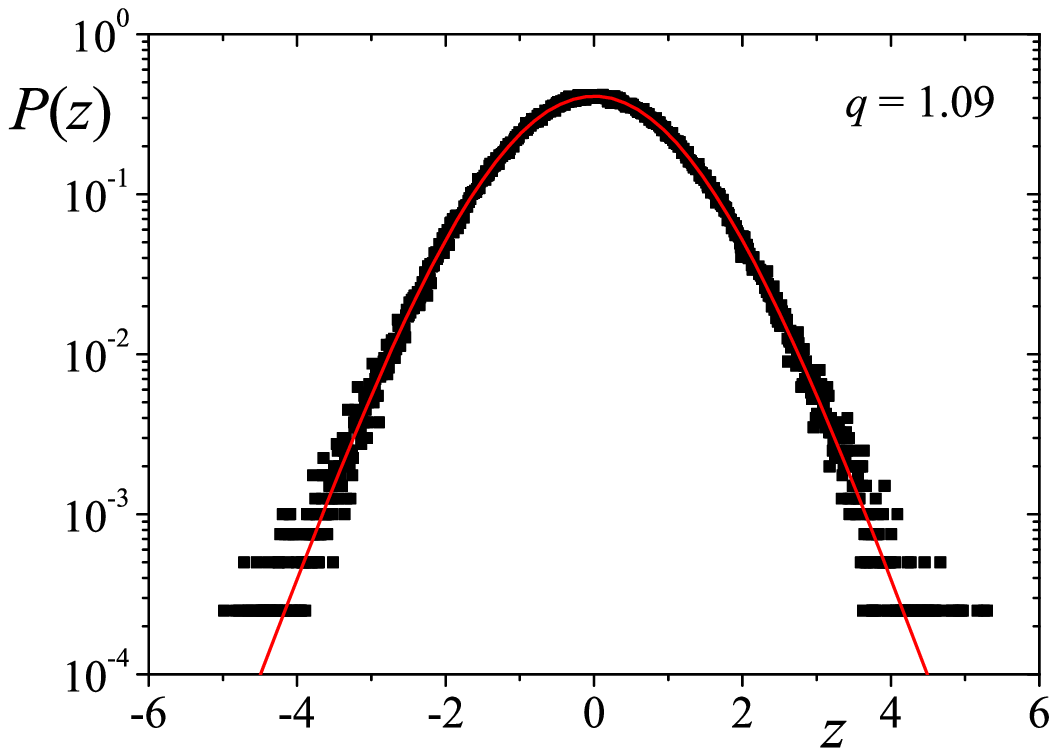} %
\includegraphics[width=0.45\columnwidth,angle=0]{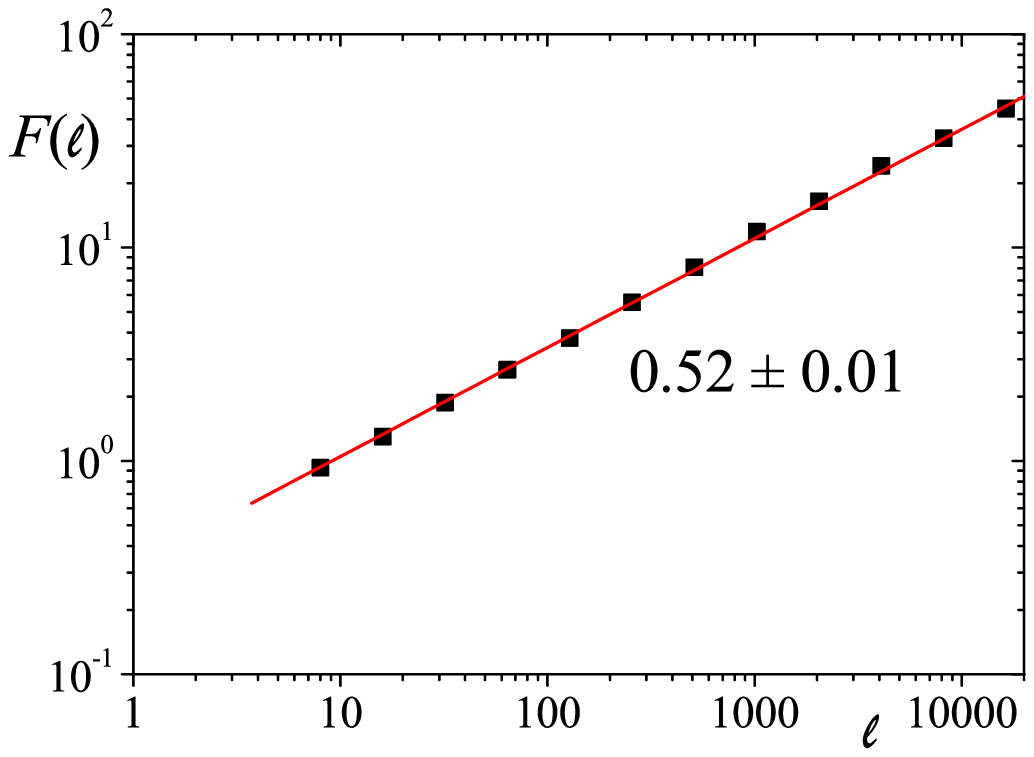} %
\includegraphics[width=0.45\columnwidth,angle=0]{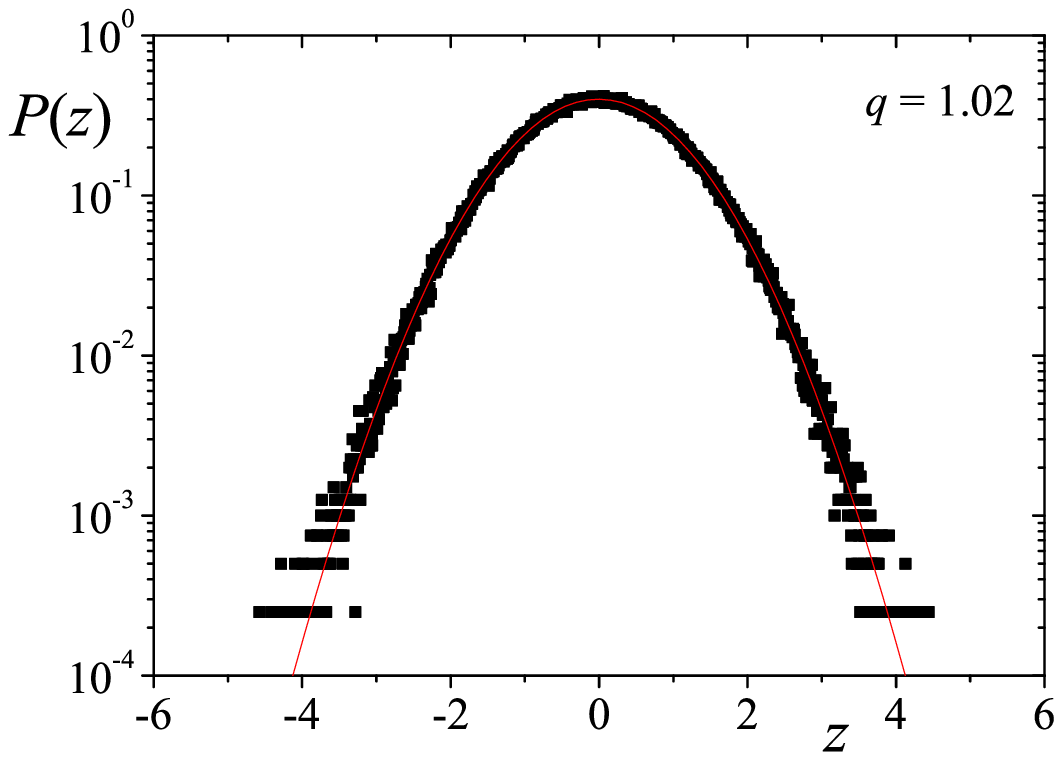} %
\includegraphics[width=0.45\columnwidth,angle=0]{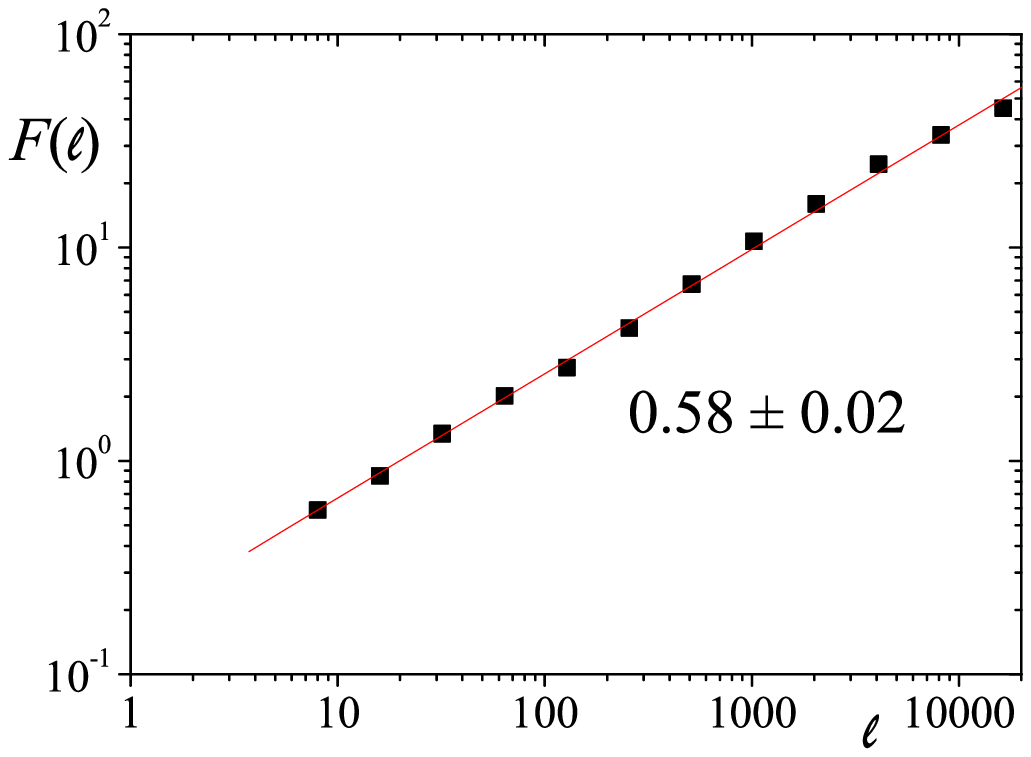}
\end{center}
\caption{Left column: Probability density functions $P(z)$ vs $z$ in a
log-linear scale; Right column: Fluctuation function $F(\ell )$ vs $\ell $
for $|z|$ in a log-log scale. The values of the model parameters are: $%
\protect\phi =0.1,W=5$ yielding $q=1$ and $H=0.5\pm 0.01$ (upper panels); $%
\protect\phi =5,W=5$ yielding $q=1.09\pm 0.01$ and $H=0.52\pm 0.01$ (middle
panels); $\protect\phi =2,W=75$ yielding $q=1.02\pm 0.01$ and $H=0.58\pm
0.02 $ (lower panels). The results have been obtained from series of $4\times
10^{5}$ elements and the numerical adjustment of $P(z)$ gave values of $%
\protect\chi ^{2}/n$ never greater than 0.00003, with $R$ never smaller than
0.998.}
\label{b05}
\end{figure}

Let us now present our results for $b=0.5$, which is able to depict the qualitative behavior of the model for small $b$. This case corresponds to a
situation of little deviation from the Gaussian, when long-range memory
is considered. In accordance, we can analyse the influence of the threshold $%
\phi $ and $W$. Overall, we verify a very sparse deviation from the Gaussian.
Keeping $W$ fixed and varying $\phi $, we understand that for small values
of $\phi $ the distribution of $z_{t}$ is Gaussian and the Hurst exponent of
$|z_{t}|$ is $1/2$. It is not hard to grasp this observation if we take into
account that, by using small values of $\phi $, we are basically employing
almost all of the past values which limits the values of instantaneous
volatility to a constant value after a transient time. As we increase the
value of $\phi $, we let the dynamics be more flexible and therefore the
volatility is able to fluctuate, resulting in a kurtosis excess. For small
values of $W$, the Hurst exponent is slenderly different from $1/2$ and the value of the Hurst exponent increases with $W$.
However, because of the small
value of $b$, the rise of $W$ turns out the distribution of $z$ barely undistinguishable from a Gaussian. This behaviour is described in
Fig.~\ref{b05}. We have obtained a Gaussian distribution and a Hurst exponent $H=0.5$ for small values of $\phi $ ($\phi = 0.1$) and $W$ ($W=5$).
When we augment the value of the threshold, $\phi =5$, the system is loose and the instantaneous volatility
is able to fluctuate leading to the emergence of tails ($q=1.09$) and a
subtle increase of the Hurst exponent ($H=0.52 \pm 0.01$). Hiking up both $W$ and $\phi
$ ($W=75$ and $\phi =2$), we have achieved large values of the Hurst
exponent ($H=0.58 \pm 0.02$), but the small value of $b$ is not sufficient to induce
relevant fluctuations, bringing on a distribution that is almost Gaussian ($%
q=1.02$). The distribution fittings were assessed by computing the critical value $\textrm{P}_{KS}^{*}=1-\alpha _{crit}$ from the Kolmogorov-Smirnov test~\cite{kolmogorov} that are equal to $0.9634$ and $0.9454$, respectively.

\begin{figure}[tbh]
\begin{center}
\includegraphics[width=0.7\columnwidth,angle=0]{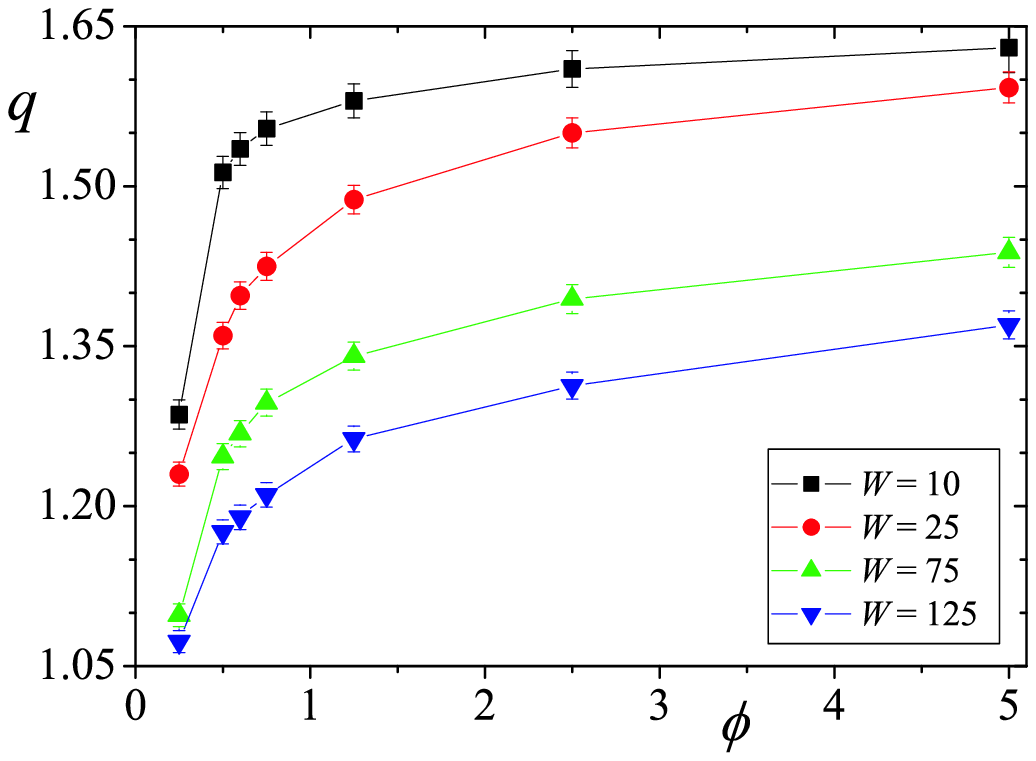} %
\includegraphics[width=0.7\columnwidth,angle=0]{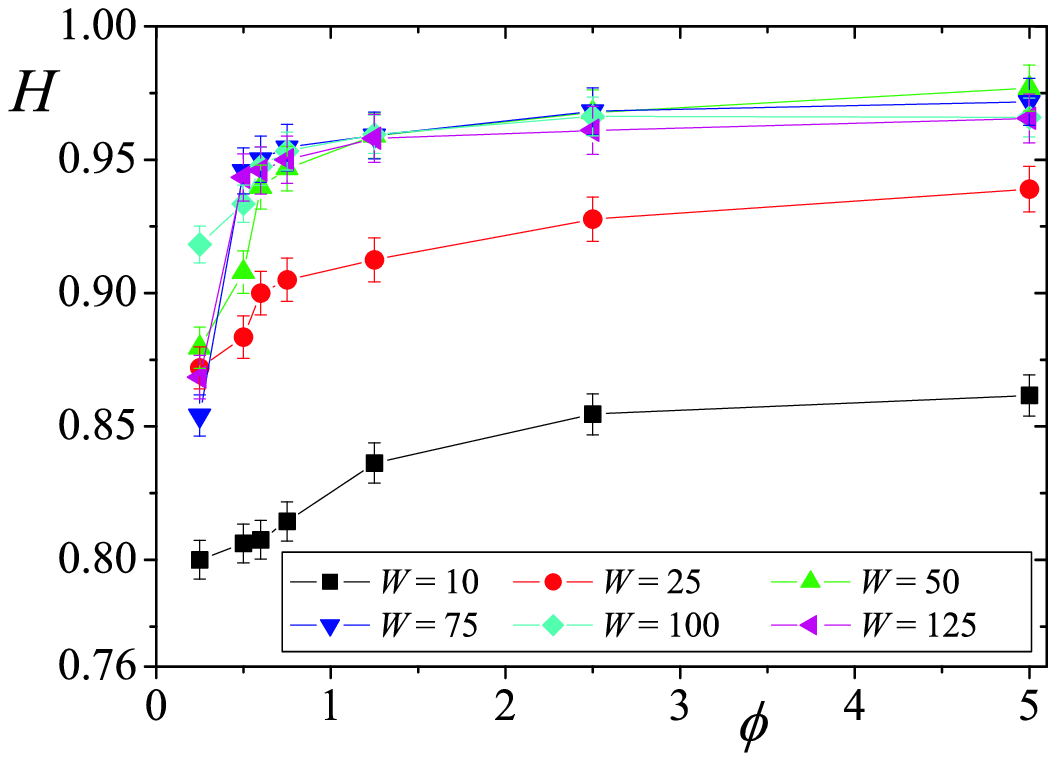}
\end{center}
\caption{Upper panel: Value of the tail index $q$ vs parameter $\protect%
\phi $ for several values of $W$ and $b=0.998$ according to the
adjustment procedures mentioned in the text. Lower panel: Hurst exponent $H$
vs $\protect\phi $. The results have been obtained from series of $4\times
10^{5}$ elements and the numerical adjustment of $P(z)$ gave values of $%
\protect\chi ^{2}/n$ never greater than 0.00003 with $R^{2}$ never smaller
than 0.9998. Regarding the values of the Hurst exponent, the absolute error
has never been greater that $0.015$ and a linear coefficient $R>0.999$.}
\label{bB}
\end{figure}

As we increase the value of $b$, we favour the contribution of the past
values of the price dynamics, thus, for the same value of $W$ we are capable
of achieving larger values of the kurtosis excess, that we represent by means
of the increase of the $q$ index. The same occurs for the Hurst exponent.
This general scenery is illustrated in Fig.~\ref{bB} for the value $b=0.998,$
where we present the dependence of $q$ and $H$ with $\phi $, for different choices of $W$.
Again, the higher $W$, the lower the tail index $q$, because the
extension of the memory surges a weakening of the fluctuations in the
volatility. The opposite occurs with the Hurst exponent, which increases towards unit (ballistic regime) as
we consider $W$ larger, for obvious reasons. In all the cases of $\left( b,W\right) $ investigated, we verified
that both $q$ and $H$ augment with $\phi $. The assessment of the numerical
adjustments is provided in Tab.~\ref{tabks} in the form of the $\textrm{P}_{KS}^{*}$ critical values
from the Kolmogorov-Smirnov test~\cite{kolmogorov}. The only case we obtained
a value $1$ (within a five-digit precision)
was for the pair $W=10$ and $\phi =5$, which results in a value
quite close to the limit of finite second-order moment (a fat-tailed
distribution with $q=5/3$). At this point it is worth saying that we have
investigated the likelihood of other well-known continuous distributions,
such as the stretched-exponential, the simple $t$-Student, L\'evy, and Gaussian. Nonetheless, the fittings carried with Eq.~(\ref{pdfz})
outperformed every other analysed distribution.

\begin{figure}[tbh]
\begin{center}
\includegraphics[width=0.6\columnwidth,angle=0]{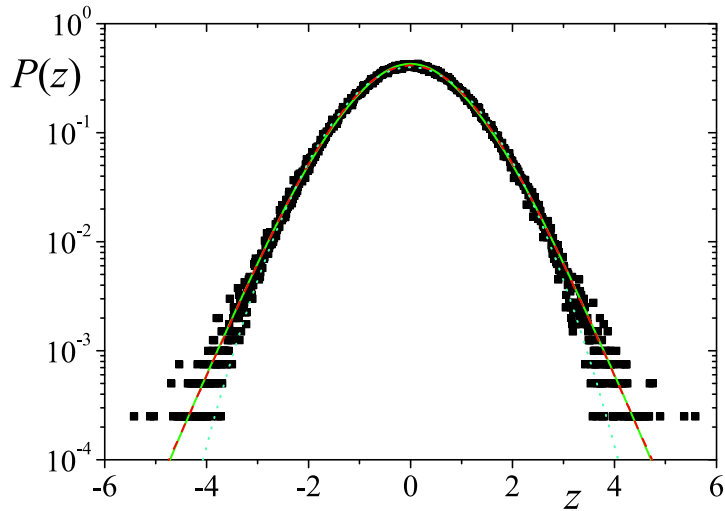} %
\end{center}
\caption{The points represent the empirical distribution function for
$b=0.998$, $\phi = 0.25$ and $W=75$; the dashed red line is our adjustment
with Eq.~(\ref{pdfz}) with $q=1.1 \pm 0.01$, $\nu = 1$ and $B=
(q-1)/(5-3\,q)$ [$\protect\chi ^{2}/n=0.00003$ and $R^{2}=0.9986$]; the
green line is PDF~(\ref{pdfuniform}) with $f=c=1/2$ and the dotted cyan line
is the Normal distribution.}
\label{uniform}
\end{figure}

\begin{figure}[tbh]
\begin{center}
\includegraphics[width=0.7\columnwidth,angle=0]{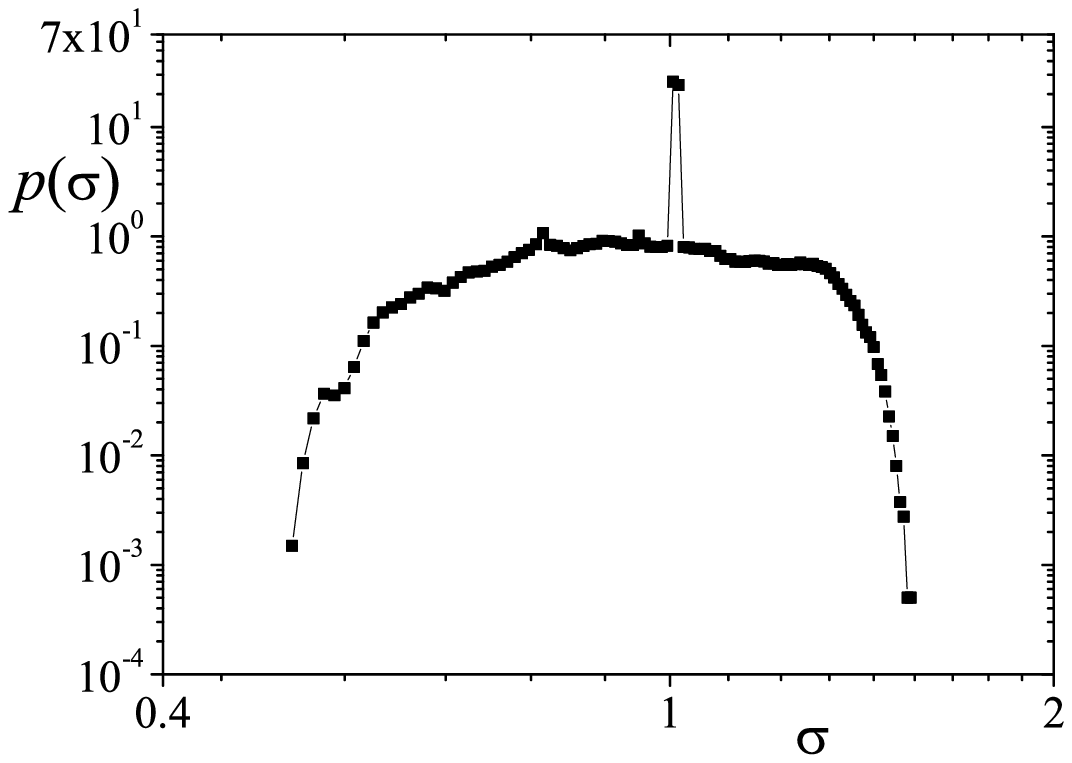} %
\includegraphics[width=0.7\columnwidth,angle=0]{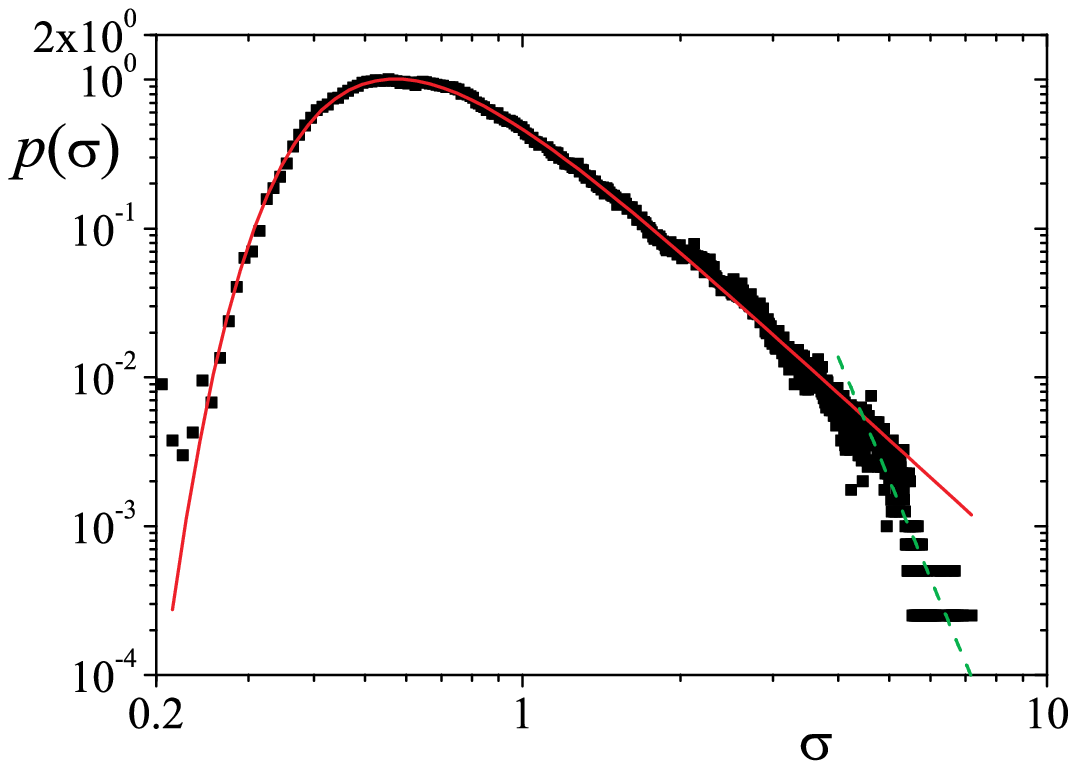}
\end{center}
\caption{Probability density function of the instantaneous volatility $p(\sigma )$ vs $\sigma $ for two different cases with $b=0.9998$.
Upper panel: $\phi = 0.25$ and $W = 75$ which leads to a sharply peaked distribution around $\sigma = 1$ and to a $P(z)$ tail index $q = 1.1$.
Lower panel: $\phi = 2.5$ and $W = 25$ that results in a broader distribution largely described by a type-2 Gumbel distribution with
$\beta = 0.421 \pm 0.002$ and $\zeta = 2.323 \pm 0.006$ ($\protect\chi ^{2}/n = 0.00011$ and $R^{2} = 0.9982$). For $\sigma \sim 5$,
$p(\sigma )$ changes its behavior to a faster decay with an exponent equal to $8.4 \pm 0.2$
represented by the gray symbols. The ANOVA test of the type-2 Gumbel
adjustment (up to $\sigma \sim 5$) have yielded a sum of squares of
$0.03553$
($323$ degrees of freedom) and $20.3684$ ($2$ degrees of freedom) for the
error and the model, respectively. The uncorrected value of the sum of
squares is $20.4039$ ($325$ degrees of freedom) and the corrected total is
$12.5941$ ($324$ degrees of freedom). The empirical distribution function
has been obtained from series of $4\times 10^{5}$ elements.}
\label{fig-vol}
\end{figure}

Concerning the instantaneous volatility, $\sigma _{t}$, we verified that
the Dirac delta distribution, $p\left( \sigma \right) =\delta \left( \sigma
-1\right) $, starts misshaping and short tails appear as we depict in Fig. %
\ref{fig-vol} (upper panel) for the case $b=0.998$, $W=75$ and $\phi =0.25$.
Considering this particular case, we can present relevant
evidence of the
effectiveness of our proposed probability distribution approach.
The empirical distribution function in the upper panel of
Fig.~\ref{fig-vol} may be simply approximated by%
\begin{equation}
p\left( \sigma \right) =\left\{
\begin{array}{ccc}
f\frac{1}{2\,c} & \mathrm{if} & \sigma \neq 1 \\
&  &  \\
\left( 1-f\right) \delta \left( \sigma -1\right)  &  & \mathrm{otherwise}%
\end{array}%
\right. ,
\end{equation}%
with $c \geq 0$, $f \leq 1$, and
$\sigma \in \left( 1-c,1+c\right)$; when $f=0$ we recover
the homoscedastic process distribution as a particular case.
Reminding that at each time step the distribution is a Gaussian (conditioned to a time-dependent value
of $\sigma $) the long-term distribution is,%
\begin{equation}
P\left( z\right) =\int_{1-c}^{1+c}p\left( \sigma \right) \frac{1}{\sqrt{%
2\,\pi }\sigma }\exp \left[ -\frac{z^{2}}{2\,\sigma ^{2}}\right] \,d\sigma ,
\end{equation}%
which gives (Ref.~\cite{fouriertransforms},~sec.~3.351),
\begin{equation}
P\left( z\right) =\frac{f}{4\sqrt{2\,\pi }\,c}\left( \mathrm{Ei}\left[ -\frac{%
z^{2}}{2\left( 1-c\right) ^{2}}\right] - \mathrm{Ei}\left[ -\frac{z^{2}}{%
2\left( 1+c\right) ^{2}}\right] \right) +\frac{1-f}{\sqrt{2\,\pi }}\exp %
\left[ -\frac{z^{2}}{2}\right]
\label{pdfuniform}
\end{equation}%
where $\mathrm{Ei}\left[ .\right] $ is the Exponential Integral function (see
e.g. Ref.~\cite{functionmeijer}). Considering $c=1/2$ (which
is appropriate to the case shown) and
taking for the sake of simplicity $f=1/2$, we obtain the
function presented
in Fig.~\ref{uniform}\footnote{%
Actually, this curve is represented in the
scaled variable $z/\sigma $
so that the standard deviation, which is originally equal to $\frac{\left(
c+1\right) ^{3}+\left( c-1\right) ^{3}+6c}{12c}$,
becomes equal to one, like in other
depicted distributions.}, the kurtosis of which is $\kappa =\frac{10854}{3125}\approx 3.47$
(making use of Ref.~\cite{fouriertransforms},~sec.~5.221).
The accordance between
this distribution and the empirical distribution is quite remarkable since
it emerges from no numerical adjustment and can be further improved by
tuning the values of $f$ and $c$. Regardless,
this kurtosis value is only $2.2\%$
larger than our numerical adjustment (see Table~\ref{tabks} for the goodness of fitting).
Furthermore, comparing the distributions by means of the symmetrised Kullback-Leibler
divergence $KL=\frac{1}{2}\left( \int P\left( z\right) \ln \frac{P\left(
z\right) }{P^{\prime }\left( z\right) }dz+\int P^{\prime }\left( z\right)
\ln \frac{P^{\prime }\left( z\right) }{P\left( z\right) }dz\right) $, we
obtain a value of 0.00014 that is 19 times smaller than the distance
between our fitting and a Gaussian. These results show that
the PDF of Eq.~(\ref{pdfz}) not only provides a good description of the data, but it is much more manageable
as well.

Cases for which the kurtosis excess is relevant ($q>5/4$) stem from
wider distributions of $\sigma $ (see the lower panel of Fig.~\ref{fig-vol}). Actually, it is the emergence of larger values of the
instantaneous volatility that brings forth fat tails. Although we have not
been successful in describing the whole distribution, we have verified that, for values of $q>5/4$,
the distribution $p\left( \sigma \right) $ is very
well described by a type-2 Gumbel distribution,%
\begin{equation}
p\left( \sigma \right) \varpropto \exp \left[ -\beta \,\sigma
^{-\zeta }\right] \,\sigma ^{-\zeta -1},  \label{Gumbel}
\end{equation}%
and after certain value of $\sigma $ the distribution sharply decreases according to a power-law with a large exponent.
We credit this sheer fall to the threshold $\phi $, which introduces a sharp change in the dynamical regime of the volatility and thus in its statistics.

In finance, such a cut-off is more than plausible as real markets do suspend trading when large price fluctuations
occur. This also grants feasibility to descriptions based on truncated power-law distributions~\cite{mantegnastanley}.
Moreover, a fall off is also presented in the quantity $\sigma _e $ of Fig.~3 in Ref.~\cite{borlandbouchaud}.
It is known that for heteroscedastic models the tail behavior of the
long-term distribution is governed by the asymptotic limit
of $p(\sigma )$ when
$\sigma $ tends to infinity. For the case of distribution~(\ref{Gumbel}),
this limit is the power-law $\sigma ^{-\zeta -1}$ and therefore we can
verify that the asymptotic behaviour of the long-term distribution of the
variable $z$,

\begin{eqnarray}
\lim_{|z|\rightarrow \infty }P\left( z\right)  &\sim &\int \left[
\lim_{\sigma \rightarrow \infty }\,p\left( \sigma \right) \right] \frac{1}{%
\sqrt{2\,\pi }\sigma }\exp \left[ -\frac{z^{2}}{2\,\sigma ^{2}}\right]
\,d\sigma  \\
&\sim &\int \sigma ^{-\zeta -2}\exp \left[ -\frac{z^{2}}{2\,\sigma ^{2}}%
\right] \,d\sigma ,  \nonumber
\end{eqnarray}
yields a power-law distribution (applying Ref.~\cite{fouriertransforms},~sec. 3.326),%
\begin{equation}
P\left( z\right) \rightarrow |z|^{-\zeta },\qquad \qquad \left(
z\rightarrow \infty \right) .
\end{equation}%
For $p(\sigma )$ following an exponential decay in the form
exp$\left[ -\gamma \,\sigma \right] $, a similar procedure
yields,%
\begin{equation}
P\left( z\right) \rightarrow G\left[ \frac{\gamma ^{2}}{8}z^{2}|%
\begin{array}{ccc}
- & - & - \\
0, & \frac{1}{2}, & 1%
\end{array}%
\right] ,\qquad \qquad \left( z\rightarrow \infty \right) ,
\end{equation}%
where $G\left[ .\right] $ is the Meijer G-function~\cite{fouriertransforms,functionmeijer}.\footnote{%
In an effort to obtain a full description of $p\left( \sigma \right) $ we also used
a function such as $f\left( x\right) =Z\exp \left[ -\beta \,x^{-\,\zeta }%
\right] \left( 1-\frac{A}{B}+\frac{A}{B}\exp \left[ \frac{A}{\mu }x\right]
\right) ^{-\,\mu }$ which allows the appearance of a crossover from a power
law to an exponential decay. Nonetheless, it did not provide better results.}

It is worth saying that we can reduce the number of parameters to $a$, $b$ and $\phi $, {\it i.e.}, apply the simple $ARCH(1)$ process,
and obtain fat tails and persistence still.

\subsection{Comparison with a real system}

Following this picture, we can now look for a set of parameters that enable
us to replicate a historic series such as the daily (adjusted\footnote{%
The adjusted values of the index take into account dividend payments and
splits occurred in a particular day.}) log-index fluctuations, $\left\{
r\left( t\right) \right\} $, of the SP500 stock index, $\left\{ S\left(
t\right) \right\} $, between 3rd January 1950 and 12th April 2010 (14380 data points) with,%
\begin{equation}
r\left( t\right) =\ln S\left( t+1\right) -\ln S\left( t\right) .
\end{equation}%
Inspecting over a grid of values of $b$, $W$ and $\phi $, we have noted that
the values of $0.9998$, $22$ and $1.125$, respectively, yield values of $q$
and $H$ for $\left\{ z_{t}\right\} $ that are in good agreement with a prior
analysis of $\left\{ r\left( t\right) \right\} $ which gave $q=1.48\pm 0.02$ (using a simple $t$-Student distribution) and
$q=1.51\pm 0.02$ $\left( q^{\prime }=1.47\pm 0.003,\nu =0.92\pm
0.008\right) $ [$\chi ^{2}/n=0.00003$, $R^{2}=0.999$ and $\textrm{P}_{KS}^{*}=0.9276$](using
the PDF of Eq.~(\ref{pdfz})) and persistence exponent
$H=0.86\pm 0.03$ (see Fig.~4). Comparing the numerical distribution of our
model with the data we obtained $D_{KS}=0.014$ and a $\textrm{P}_{KS}^{*}$ critical value equal to
$0.991$ from the two-sample Kolmogorov-Smirnov
test~\cite{kolmogorov},while
the comparison between the distribution of the numerical procedure and the
adjustment of the SP500 empirical distribution function yielded $\textrm{P}_{KS}^{*}=0.9998$.
Once again we have tested other possible numerical adjustments
and the only other
relevant distribution was the stretched exponential with $\nu =1.3 \pm
0.02 $ $(q ^{\prime } = 1)$ which has given a $\textrm{P}_{KS}^{*}$ different from $1$ $\left(
\textrm{P}_{KS}^{*}=0.9999\right)$, but a significantly larger value of $\chi ^{2} $
[$\chi ^{2}/n=0.00009$, $R^{2}=0.9963$].

It is worthy to be mentioned that
all the three values of the parameters are plausible. First, within an application context, $b$
is traditionally a value robustly greater than $0.9$. Second, $W$ is close
to the number of business days in a month and last, but not least, $\phi $
is somewhat above the average level of the mean variance presented above.
This provides us with a very interesting picture of the dynamics.
Specifically, at a relevant approximation we can describe this particular
system as monitoring the magnitude of its past fluctuations with a
characteristic scale of a month, from which it computes the level of impact
resulting in an excess of volatility. Actually one month moving averages are established indicators in quantitative analyses of financial markets.
When the volatility in a period of the
same order of magnitude of $W$ surpasses the value $\phi /\left( 1-b\right) $,
then the system recalls previous periods of time, no matter how
long they happened, in which a significant level of volatility excess
occurred. Those periods are then averaged in order to determine the
level of instantaneous volatility $\sigma _{t}^{2}$.

\begin{figure}[tbh]
\begin{center}
\includegraphics[width=0.45\columnwidth,angle=0]{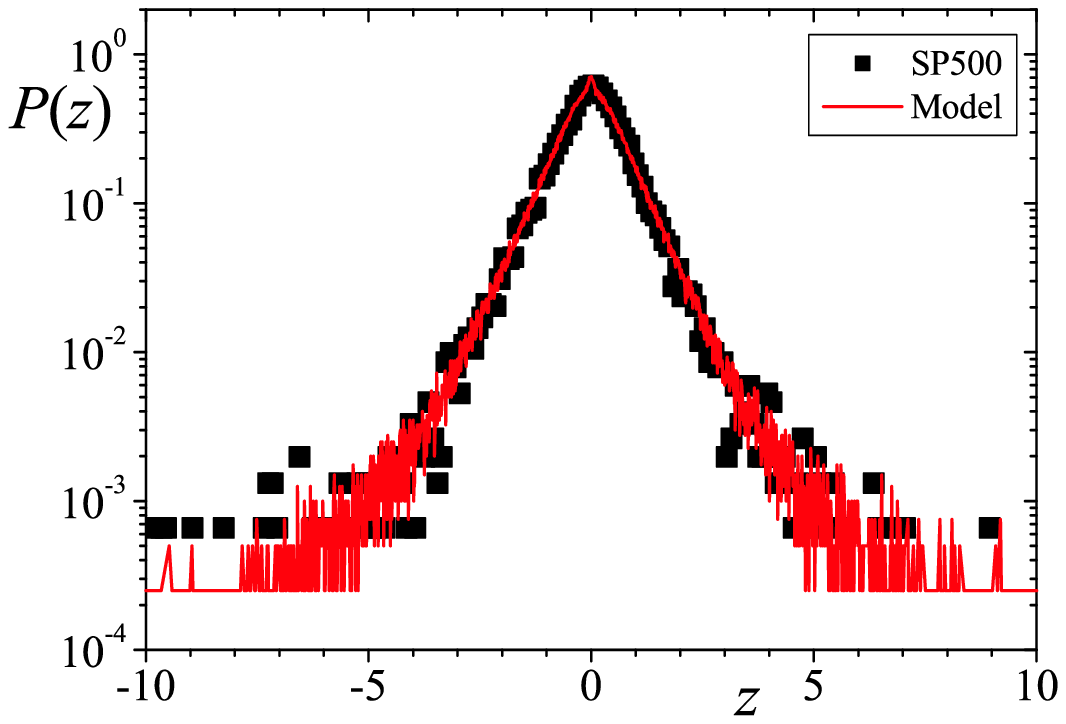} %
\includegraphics[width=0.45\columnwidth,angle=0]{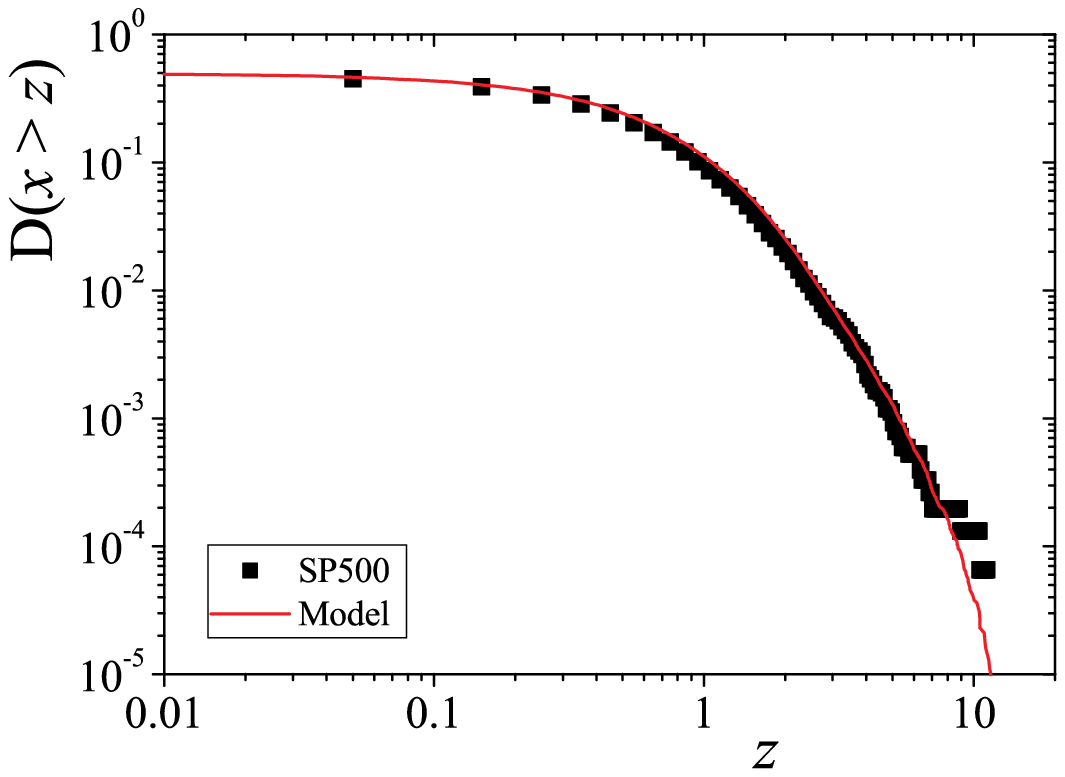} %
\includegraphics[width=0.45\columnwidth,angle=0]{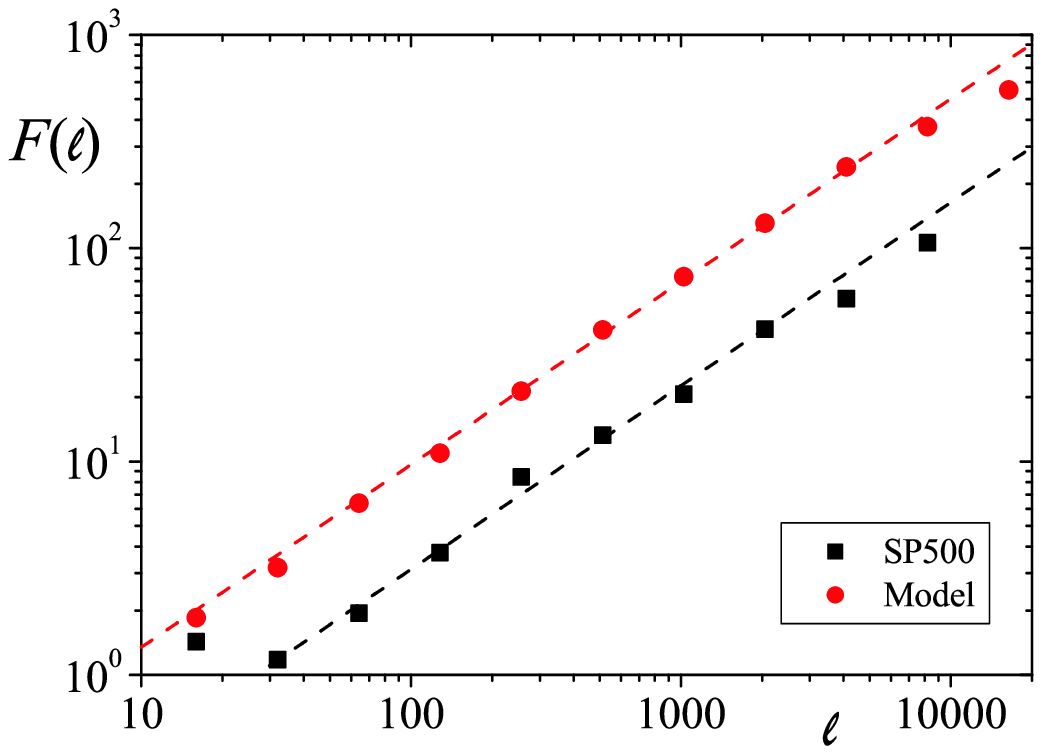}
\end{center}
\label{bSP}
\caption{Upper panels: On the left side, Probability density function $P(z)$ vs $z$ for $%
b=0.998$, $W=22$ and $\protect\phi =1.125$ (full line) [$q=1.49\pm 0.01$
with $\protect\chi ^{2}/n=0.000025$ and $R^{2}=0.9984$] and the $SP500$
daily log-index fluctuations (symbols) [$q=1.48\pm 0.02$ with $\protect\chi %
^{2}/n=0.00004$ and $R^{2}=0.996$] in the log-linear scale and on the right side the complementary cumulative
distribution function $D(x > z)$ vs $z$ for case shown on the left. Lower panel:
Fluctuation function $F(\protect\ell )$ vs $\protect\ell $ for the same parameters above
[$H=0.85\pm 0.02$, with $R=0.998$] (red circles) and for the $SP500$ daily log-index fluctuations [$%
H=0.86\pm 0.03$, with $R=0.997$] (black squares) in a log-log scale.}
\end{figure}

\section{Discussion}

We have studied a generalisation of the well-known $ARCH$ process born in a financial context.
Our proposal differs from other generalisations, since it adds to heteroscedastic dynamics the ability to reproduce
systems where cognitive traits exist or systems showing typical cut-off limiting values. In the former case, when present circumstances are close to extreme and impacting events, the dynamics switches to the memory of abnormal events. By poring over the set of parameters of the problem,
namely the impact of past values, $b$, the memory scale, $W$, and the volatility threshold, $\phi$, we have verified that we are
able to obtain times series showing fat tails for the probability density function and strong persistence for the magnitudes of
the stochastic variable (directly related to the instantaneous volatility), as it happens in several
processes studied within the context of complexity. In order to describe the usefulness of our model we have applied it to mimic the fluctuations of the stock index $SP500$, we verified that the best values reproducing the features
of its time series are $W$ close to
one business month and $\phi$ greater that the mean variance of the process which is much larger than the normal level
of volatility for which trading is not taken into account.
Concerning the volatility, we have noticed that for the problems of interest (\emph{i.e.}, fat tails and strong persistence),
the distributions are very well described by a type-2 Gumbel distribution in large part of the domain, which explains the emergence of the tails.

\ack{
SMDQ thanks the warm hospitality of the CBPF and its staff during his visits to the institution sponsored by CNPq and National Institute of Science and Technology for Complex Systems and the financial support of the European Commission through the Marie Curie Actions FP7-PEOPLE-2009-IEF (contract nr 250589) in the final part of the work. EMFC and FDN acknowledge the financial support of CNPq and FAPERJ and CNPq, respetively.
The current version of our work benefited from the comments of incognitos peers to whom we are grateful.
}

\end{document}